\documentclass[Physsubmission, Phys]{SciPost}

\hypersetup{
    colorlinks,
    linkcolor={red!50!black},
    citecolor={blue!50!black},
    urlcolor={blue!80!black}
}

\usepackage[bitstream-charter]{mathdesign}
\usepackage{graphicx}
\usepackage[utf8]{inputenc}
\usepackage{booktabs}
\usepackage{hyperref}
\usepackage[font={small}]{caption}
\usepackage[font={small}]{subcaption}
\usepackage{color}
\usepackage{mathtools}
\usepackage{float}
\usepackage[nottoc, notlot, notlof]{tocbibind}
\usepackage{physics}
\usepackage{bigints}
\usepackage[makeroom]{cancel}
\usepackage{braket}
\usepackage{geometry}

\DeclareSymbolFont{usualmathcal}{OMS}{cmsy}{m}{n}
\DeclareSymbolFontAlphabet{\mathcal}{usualmathcal}

\begin{document}

\begin{center}{\Large \textbf{
$N^3LO$ calculations for $2\to 2$ processes using Simplified Differential Equations\\
}}\end{center}

\begin{center}
Dhimiter D. Canko\textsuperscript{1, 2 $\star$},
Federico Gasparotto\textsuperscript{3, 4},
Luca Mattiazzi\textsuperscript{3, 4},
Costas G. Papadopoulos\textsuperscript{2} and
Nikolaos Syrrakos\textsuperscript{2, 5}
\end{center}

\begin{center}
{\bf 1} Department of Physics, University of Athens, Zographou 15784, Greece
\\
{\bf 2} 
Institute of Nuclear and Particle Physics, NCSR “Demokritos”
Agia Paraskevi 15310, Greece
\\
{\bf 3} Dipartimento di Fisica e Astronomia, Università di Padova, Padova 35131, Italy
\\
{\bf 4} INFN, Sezione di Padova, Padova 35131, Italy
\\
{\bf 5} Physics Division, National Technical University of Athens, Athens 15780, Greece
\\
* jimcanko@phys.uoa.gr
\end{center}

\begin{center}
\today
\end{center}


\definecolor{palegray}{gray}{0.95}
\begin{center}
\colorbox{palegray}{
  \begin{tabular}{rr}
  \begin{minipage}{0.1\textwidth}
    \includegraphics[width=35mm]{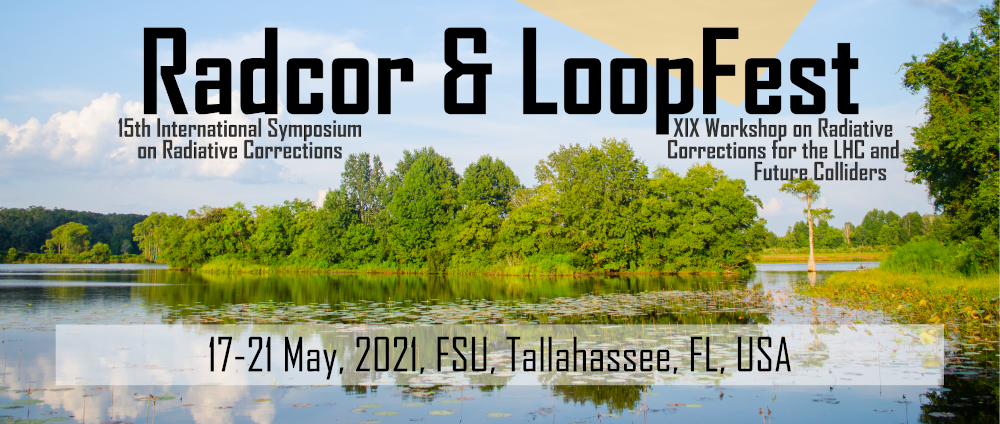}
  \end{minipage}
  &
  \begin{minipage}{0.85\textwidth}
    \begin{center}
    {\it 15th International Symposium on Radiative Corrections: \\Applications of Quantum Field Theory to Phenomenology,}\\
    {\it FSU, Tallahasse, FL, USA, 17-21 May 2021} \\
    \doi{10.21468/SciPostPhysProc.?}\\
    \end{center}
  \end{minipage}
\end{tabular}
}
\end{center}

\section*{Abstract}
{\bf

We present the computation of the massless three-loop ladder-box family with one external off-shell leg using the Simplified Differential Equations (SDE) approach. We also discuss the methods we used for finding a canonical differential equation for the two tennis-court families with one off-shell leg, and the application of the SDE approach on these two families.}

\vspace{10pt}
\noindent\rule{\textwidth}{1pt}
\tableofcontents\thispagestyle{fancy}
\noindent\rule{\textwidth}{1pt}
\vspace{10pt}

\section{Introduction}
\label{sec:intro}

We are living in very interesting times to be a particle physicist. The ever-increasing accuracy of the experimental measurements and the future runs of LHC, HL-LHC and new collider experiments will demand the most precise theoretical predictions for their interpretation. Possible small deviations between experiment and theory will make apparent the existence of new phenomena and will dethrone once and for all the \textit{Standard Model}, which is already facing existential issues due to its incompatibility with astrophysical observations (\textit{Dark Matter/Energy}) and its own components (\textit{Neutrino Oscilations}). 

From the theory point of view, the high-precision predictions can be obtained using \textit{Perturbative Quantum Field Theory}. Within this framework, the current frontier for $2 \to 2$ scattering processes stands at $N^3LO$, where the computation of three-loop \textit{Feynman Integrals} (FI) is demanded. From these FI, all the families with massless internal and external particles have been calculated \cite{3loop1,3loop2,3loop3,3loop4,3loop5} and have been very recently used for the computation of 3-loop 4-point \textit{Amplitudes} (for the first time in QCD) for the processes $q\bar{q}\to \gamma \gamma$ \cite{Tancredi1} and  $q\bar{q} \to q \bar{q}$\footnote{Where the initial and the final state quarks can have different flavour.} \cite{Tancredi2}. As it regards the massless families with one external off-shell leg, which are relevant for processes like $e^{+} e^{-} \to \gamma^* \to 3j$, $pp \to Z j$ and $pp \to H j$, only one has been calculated \cite{3loop5,3loop6}, while no progress has been made on the computation of the families with two off-shell legs so far.

The modern approach for computing FI is using the method of \textit{Differential Equations} (DE) \cite{de1,de2,de3,de4}, which is utilized within the framework of \textit{Dimensional Regularization} ($d=4-2\varepsilon$) and the FI are computed as a Laurent expansion in $\varepsilon$. This method takes advantage of the fact that the FI are functions of the \textit{Mandelstam variables}, thus one can differentiate with respect to them, and that any FI of a family can be written as a linear combination of a finite basis, called \textit{Master Integrals} (MI), which is implied by the \textit{Integration-By-Parts relations} (IBP) \cite{IBP1,IBP2}. The basis of MI is not unique and a proper choice of it, in such a way such it consists MI that are \textit{Pure Functions}\footnote{In the following we will refer to this basis as UT basis} \cite{Henn1}, leads to a DE of the so-called \textit{Canonical Form} \cite{Henn1}. This DE is $\varepsilon-$factorized, \textit{Fuchsian} and the residue matrices are purely numerical, thus can be iteratively solved at any order on $\varepsilon$.  

In the next sections we use the \textit{Simplified Differential Equation} approach (SDE) \cite{SDE1,SDE3}, which is a variant of the DE method, combined with a UT basis \cite{SDE4} in order to solve the massless ladder-box family with up to one leg off-shell. Within the SDE the external momenta are parameterized in terms of $x$, a dimensionless parameter which is introduced in such a way such to capture the off-shellness of an external leg, and the DE is created by taking derivatives with respect to $x$. An extra feature of the SDE is that one can almost for free obtain the solution for the same family with one external massive leg less, by taking the $x\to 1$ limit. We also discuss the methods we used for obtaining a DE of canonical form for the two tennis-court families with one off-shell leg.

\section{Three$-$loop massless ladder-box with up to one off-shell leg}

\subsection{Massless ladder-box with one off-shell leg}

This family is described by the Feynman graph of the Figure 1 and contains a set of 83 MI, as we found using \texttt{Kira 2.0} \cite{kira} and \texttt{FIRE6} \cite{fire,litered}. In this computation we adopt the notation for the kinematics and the UT basis from \cite{3loop6}, where this family was first studied. The class of FI describing this family can be expressed via the following expression
\begin{figure}
\centering
\begin{subfigure}{.5\textwidth}
  \centering
  \includegraphics[width=1.0\textwidth]{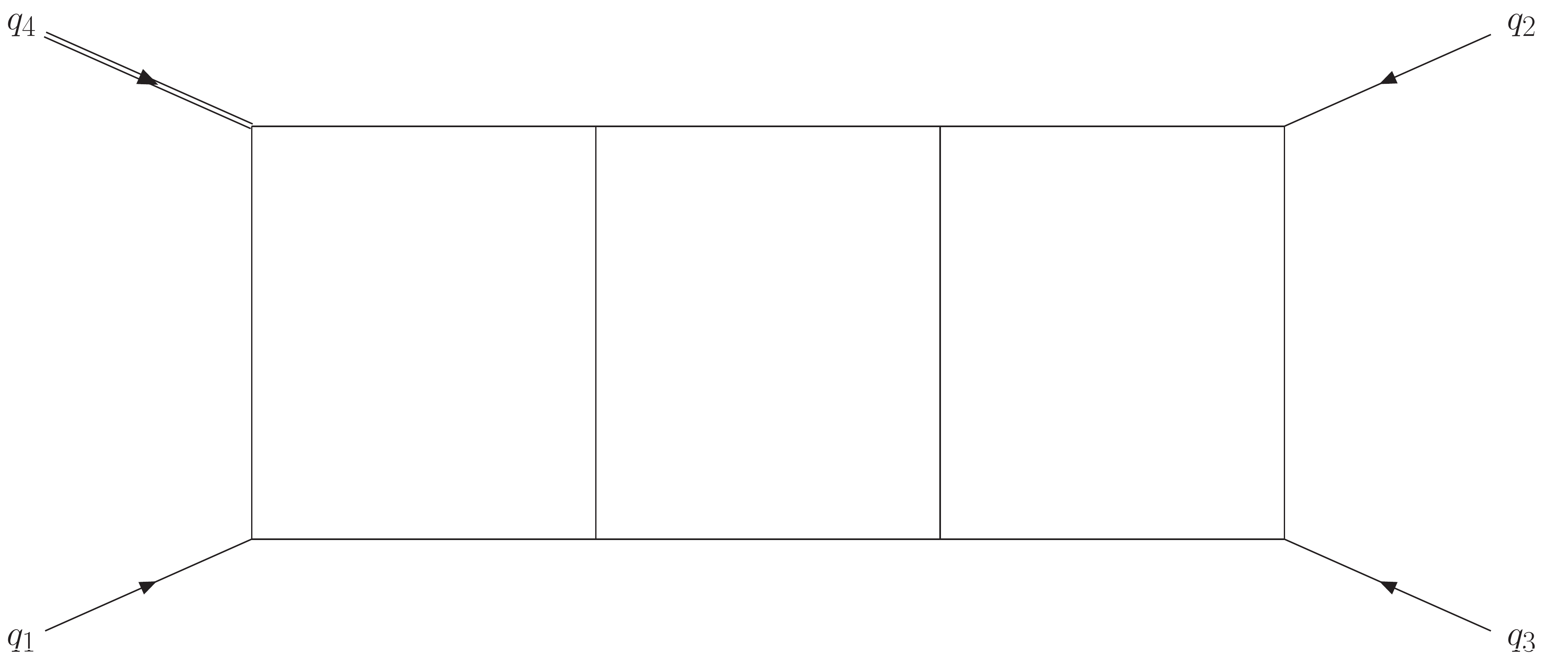}
  \caption{In the standard notation.}  
\end{subfigure}%
\begin{subfigure}{.5\textwidth}
  \centering
  \includegraphics[width=1.0\textwidth]{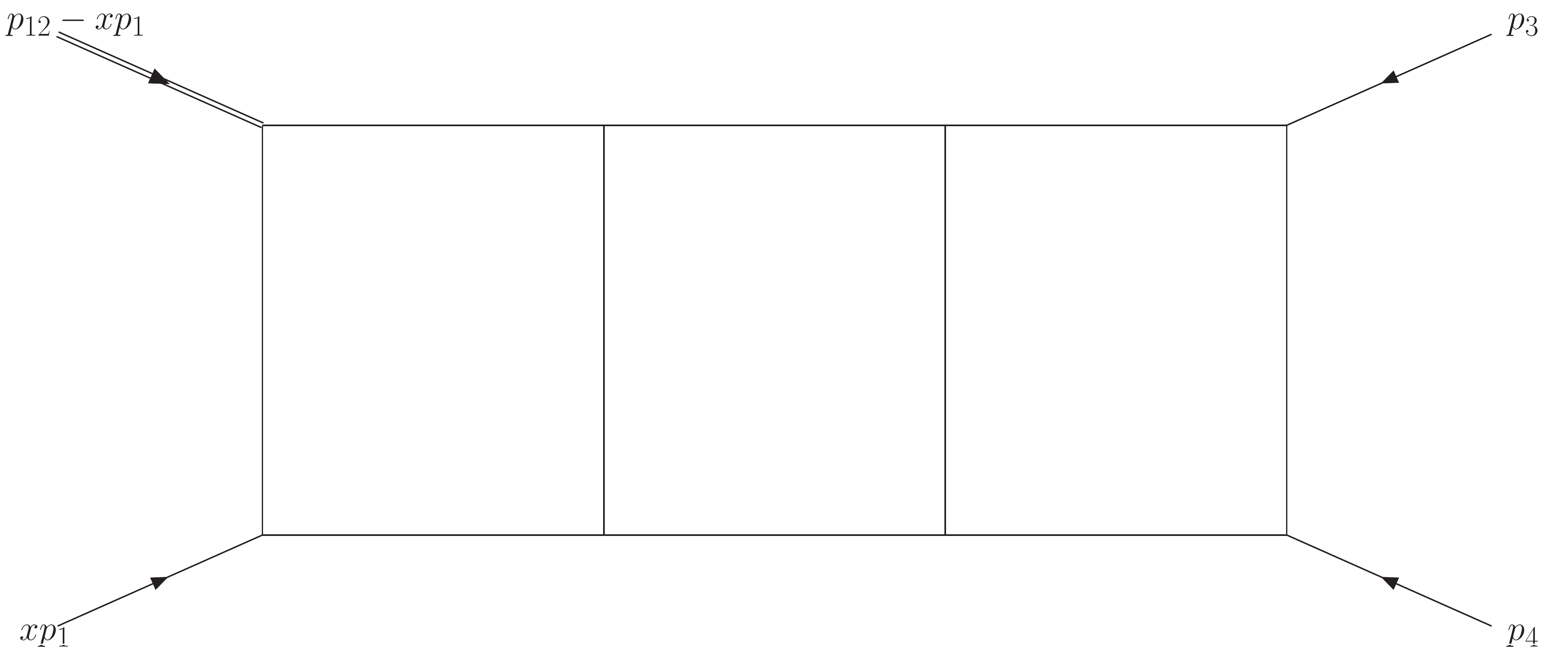}
  \caption{In the SDE notation.}
\end{subfigure}
\caption{The Feynman graph of the three-loop massless ladder-box family with one off-shell leg. All the external particles are assumed incoming.}
\end{figure}
\begin{equation}
\label{family} 
G_{a_1,\dots,a_{15}}\left( \{ q_j \}, \varepsilon \right)=\int \left( \prod_{r=1}^3 \frac{d^d l_r}{i \pi^{d/2}} \right) \frac{e^{3 \varepsilon \gamma_E}}{D_{1}^{a_1} \dots D_{15}^{a_{15}}} \quad \text{with} \quad d=4-2\varepsilon \, ,
\end{equation}
where $D_{11},\dots,D_{15}$ are propagators coming from \textit{Irreducible Scalar Products} (thus for them we have $\{a_{11},a_{12},a_{13},a_{14},a_{15}\}\leq 0$), and the chosen parametrization for the propagators is\footnote{from now on we use the abbreviation $q_{i\dots j}= q_i +\dots + q_j$ and $p_{i\dots j}=p_i+\dots+p_j$.}
\begin{equation}
\label{propagators}
\begin{split}
&D_1=l_1^2, \quad D_2=l_2^2, \quad D_3=l_3^2, \quad D_4=(l_1-l_2)^2, \quad D_5=(l_2-l_3)^2,\\
&D_6=(l_3+q_2)^2, \quad D_7=(l_1+q_{23})^2, \quad D_8=(l_2+q_{23})^2, \quad D_9=(l_3+q_{23})^2,\\
&D_{10}=(l_1+q_{123})^2, \quad D_{11}=(l_1+q_2)^2, \quad D_{12}=(l_2+q_2)^2,\\ &D_{13}=(l_2+q_{123})^2, \quad D_{14}=(l_3+q_{123})^2, \quad \text{and} \quad D_{15}=(l_1-l_3)^2.
\end{split}
\end{equation}
We define the Mandelstam variables from the external momenta ($q_1^2=q_2^2=q_3^2=0 \, \, \text{and} \, \, q_4^2=m^2$) using the notation
\begin{equation}
\label{invariants}
q_2 \cdot q_3=s/2, \quad q_1 \cdot q_3=t/2, \quad q_1 \cdot q_2=(m^2 - s - t)/2 .
\end{equation}
In order to apply the SDE approach we chose the following one $x-$parametrization
\begin{equation}
\label{SDE}
q_1 \rightarrow x p_1, \quad q_2 \rightarrow p_3, \quad q_3 \rightarrow p_{4}, \quad q_4 \rightarrow p_{12} -x p_1  \quad \text{with} \quad p_1^2=p_2^2=p_3^2=p_4^2=0,
\end{equation}
where the initial invariants of \eqref{invariants} are now parametrized in terms of $x$ and the Mandelstam variables of the on-shell momenta ($s_{12}=p_{12}^2$ and $s_{23}=p_{23}^2$)
\begin{equation}
\label{SDEinva}
s=s_{12}, \quad t=x s_{23}, \quad m^2=(1-x) s_{12},
\end{equation}
and $x$ is introduced in 3 propagators ($D_{10}, D_{13}, D_{14}$).

Differentiating with respect to $x$ and using the UT basis of \cite{3loop6} and the IBP relations we obtained a canonical DE
\begin{equation}
\label{de}
\partial_{x} \textbf{g}=\varepsilon \left( \sum_{i=1}^4 \frac{\textbf{M}_i}{x-l_i} \right) \textbf{g},
\end{equation} 
with the four letters being $l_i=\{0,\,1,\, s_{12}/(s_{12}+s_{23}),-s_{12}/s_{23}\}$. We solve the DE up to weight six on $\varepsilon$ and in the Euclidean region $\{0<x<1$, $s_{12}<0$, $s_{12}<s_{23}<0\}$, where the FI are free of branch cuts. The solution has the following form
\begin{equation}
\begin{split}
\textbf{g}&=\varepsilon^0 \textbf{b}_0^{(0)}+\varepsilon \left(\sum {\cal G}_i \textbf{M}_i \textbf{b}_0^{(0)}+\textbf{b}_0^{(1)}\right)+\varepsilon^2 \left(\sum {\cal G}_{ij} \textbf{M}_i\textbf{M}_j\textbf{b}_0^{(0)}+\sum {\cal G}_i \textbf{M}_i \textbf{b}_0^{(1)}+\textbf{b}_0^{(2)} \right)+ \dots \\  
&+ \varepsilon^6 \left(\textbf{b}_0^{(6)}+ \sum {\cal G}_{ijklmn} \textbf{M}_i \textbf{M}_j \textbf{M}_k \textbf{M}_l \textbf{M}_m \textbf{M}_n \textbf{b}_0^{(0)} + \sum {\cal G}_{ijklm} \textbf{M}_i \textbf{M}_j \textbf{M}_k \textbf{M}_l \textbf{M}_m \textbf{b}_0^{(1)} \right. \\
&+\left. \sum {\cal G}_{ijkl} \textbf{M}_i \textbf{M}_j \textbf{M}_k \textbf{M}_l \textbf{b}_0^{(2)} +\sum {\cal G}_{ijk} \textbf{M}_i\textbf{M}_j\text{M}_k \textbf{b}_0^{(3)}+\sum {\cal G}_{ij} \textbf{M}_i\textbf{M}_j\textbf{b}_0^{(4)}+\sum {\cal G}_i \textbf{M}_i \textbf{b}_0^{(5)} \right) \, ,
\end{split}
\end{equation}
where the matrices $\textbf{b}_0^{(i)}$ are the boundaries and ${\cal G}_i, ... ,{\cal G}_{ijklmn}$ are Goncharov poly-logarithms \cite{GPL} of weight $1, \dots,6$, respectively, with argument $x$ and letters from the set $l_i$. For the manipulation of these poly-logarithms we have used \texttt{PolyLogTools} \cite{PolyLogTools}.

As it regards the calculation of the boundary conditions, we start by taking advantage of the fact that some MI are already known in closed form and thus we can directly obtain boundary conditions for them. Afterwards, we use the fact that if for a basis element its leading regions contributing to its asymptotic limit \cite{asy1,asy2,asy3,asy4,fiesta} $x \to 0$ are of the form $x^{\alpha+\beta \varepsilon}$ with $\alpha\geq 1$ its boundary should vanish. Then by comparing the regions found by \texttt{asy} with that found by the resummation matrix method \footnote{For an earlier use of the Jordan decomposition method see also \cite{mas1,mas2}.} we obtain relations between different boundaries. In fact, we obtain two kinds of relations \cite{3loop5,SDE4}. The first of them we call it \textit{pure relations} because contain only boundaries of the basis elements, while the second of them we call it \textit{impure} due to the fact that are relations between boundaries and asymptotic limits. In the end, we are left with some regions to calculate in order to determine all the boundaries, which we do so by using standard expansion-by-regions techniques\footnote{Meaning calculating the hard regions in the momentum-space while the soft regions in the Feynman parameter representation.}.

We crossed-checked our results with the ones from \cite{3loop6} and we found perfect agreement for all the MI.

\subsection{Massless ladder-box}

\begin{figure} [h!]
\centering
\includegraphics[width=3.5 in]{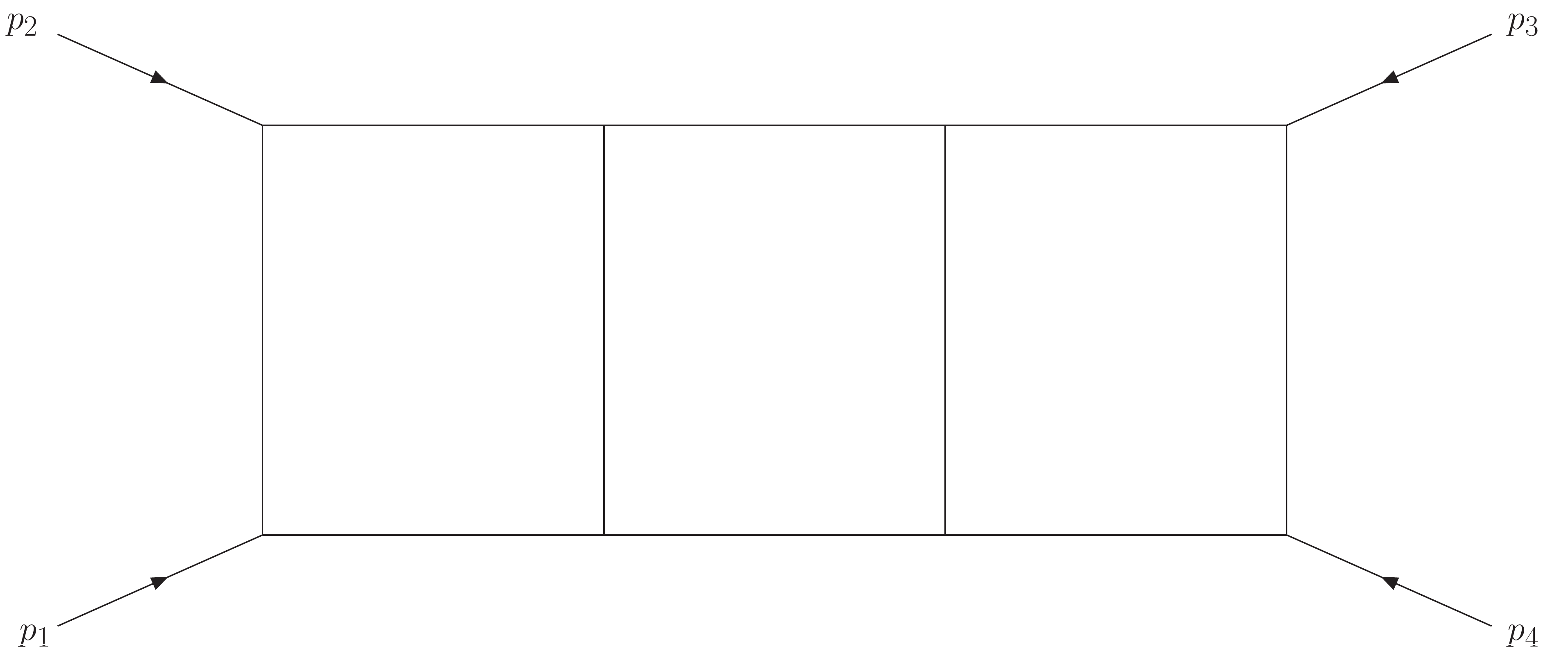}
\caption{The Feynman graph of the three-loop massless ladder-box family.}
\end{figure}

From the solution of the ladder-box with one off-shell leg, by taking the $x\to 1$ limit within the SDE \cite{3loop5, SDE3}, we also obtain the solution for a UT basis of the massless ladder-box family. The procedure for obtaining this solution is the following 
\begin{enumerate}
  \item Expand the solution in terms of $\log(1-x)$:
\begin{equation}
    \textbf{g}=\sum_{n\geq 0} \epsilon^{n} \sum_{i=0}^{n}\frac{1}{i!}\textbf{c}^{(n)}_{i} \log^i(1-x)
\end{equation}
  \item From the above expansion define the regular part of $\textbf{g}$ at $x=1$ and from it the truncated part:
\begin{equation}
   \textbf{g}_{reg}=\sum \epsilon^n \textbf{c}^{(n)}_{0} \quad \text{and} \quad \textbf{g}_{trunc} = \left.\textbf{g}_{reg}\right|_{x=1}
\end{equation}
  \item Define the resummation matrix $\textbf{R}_1$ and from it the numerical matrix $\textbf{R}_{10}$:
\begin{equation}
    \textbf{R}_1 = e^{\epsilon \textbf{M}_1 \log(1-x)} = \textbf{S}_1 e^{\epsilon \textbf{D}_1 \log(1-x)} \textbf{S}_{1}^{-1} \quad \text{and} \quad  \textbf{R}_1 \xrightarrow{(1-x)^{a_{i} \epsilon}\to 0} \textbf{R}_{10}
\end{equation}
  \item Find the $x\to1$ limit by acting $\textbf{R}_{10}$ to $\textbf{g}_{trunc}$:
\begin{equation}
    \textbf{g}_{x\to 1} = \textbf{R}_{10} \textbf{g}_{trunc}
\end{equation}
  \item Reduce the number of the basis elements to that of the MI of the massless ladder-box using the property $\textbf{R}_{10}^2=\textbf{R}_{10}$ and/or IBP.
\end{enumerate}
For the FI of this family we have chosen the following normalization
\begin{equation}
G_{a_{1}, \ldots, a_{15}}\left(\left\{p_{j}\right\}, \varepsilon\right)=(-s_{12})^{3\varepsilon}\int\left(\prod_{l=1}^{3} \frac{d^{d} k_{l}}{i \pi^{d / 2}}\right) \frac{e^{3 \varepsilon \gamma_{E}}}{D_{1}^{a_{1}} \ldots D_{15}^{a_{15}}},
\end{equation}
where the propagators are obtained by setting $x=1$ to the propagators of the massive family. We compared analytically our results with the ones given by \cite{3loop2} and numerically with pySecDec \cite{pysecdec} in the Euclidean region. In both cases, we found perfect agreement.

\section{Canonical DE construction for the two tennis-court families}

\begin{figure}
\centering
\begin{subfigure}{.5\textwidth}
  \centering
  \includegraphics[width=0.8\textwidth]{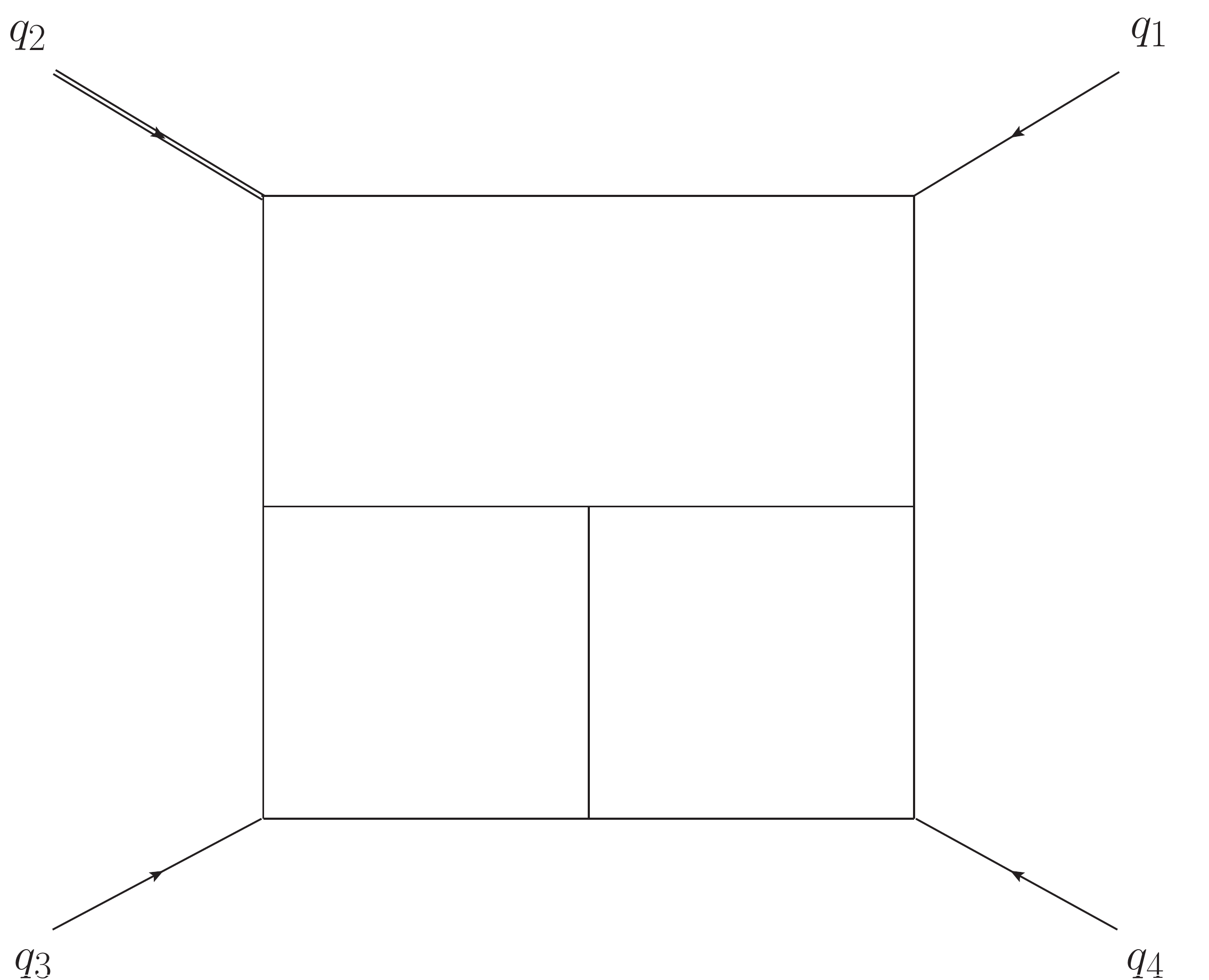} 
  \caption{The F2 family}
\end{subfigure}%
\begin{subfigure}{.5\textwidth}
  \centering
  \includegraphics[width=0.8\textwidth]{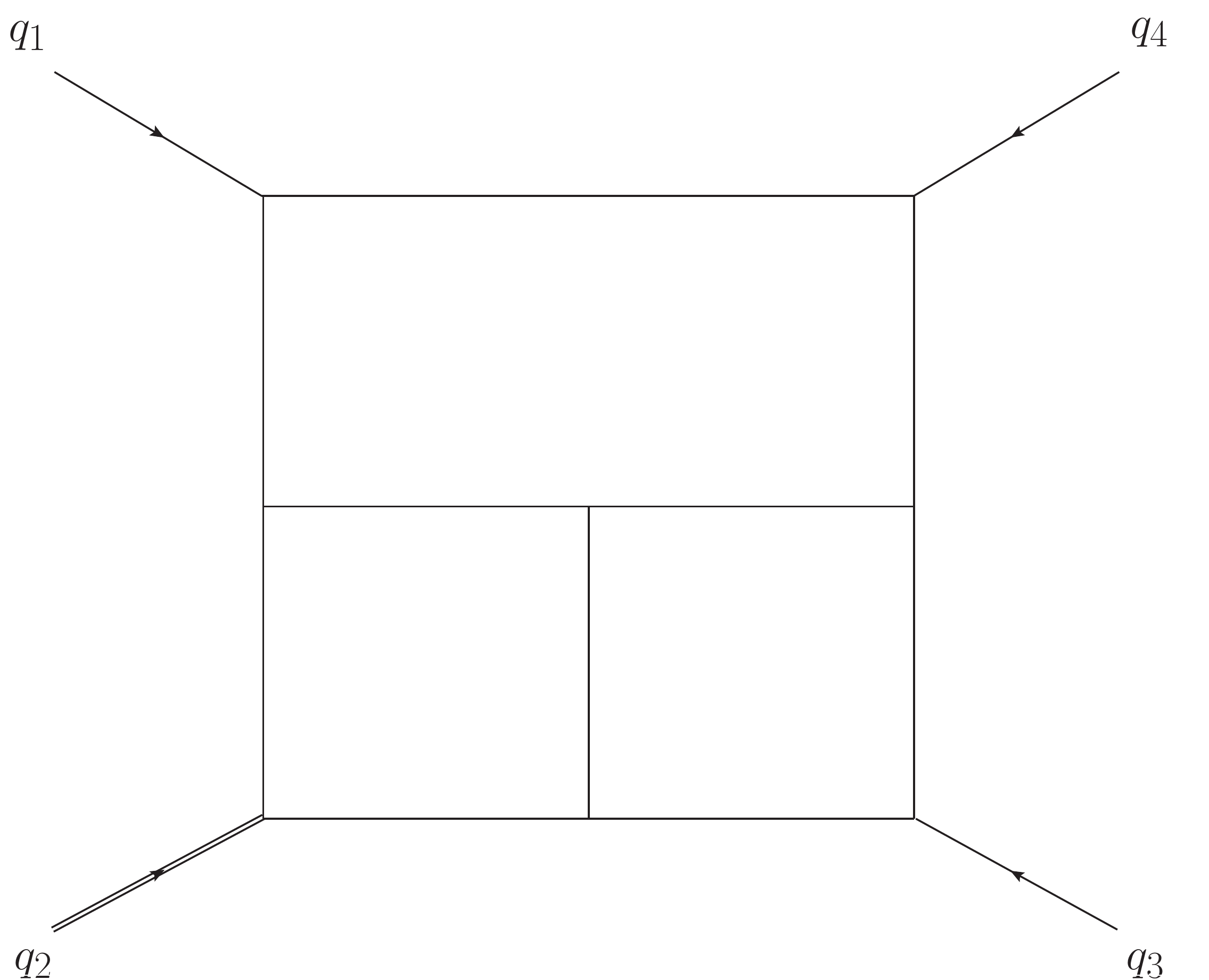}
  \caption{The F3 family}
\end{subfigure}
\caption{The Feynman graphs of the two tennis-court families.}
\end{figure}

For the completion of the computation of all the massless three-loop four-point planar MI with one external massive particle, apart from the ladder-box family one also needs to calculate the MI of the two massless tennis-court families with one off-shell leg (which are depicted in Figure 3). The first of them contains 117 MI and we denote it as F2, while the second of them contains 166 MI and we denote it as F3. Both families have the same letters with the ladder-box family. In this section we briefly present the methods that led us to the construction of a UT basis for the two tennis-court families. In total (in both families) there exist 91 new MI and for finding their corresponding UT basis element we used three methods.

One of the methods is the \textit{Magnus Exponential} \cite{magnus} method applied to the DE derived by differentiating with respect to the Mandelstam variables, which we used for some lower-sector (till 7 propagators) MI\footnote{It is important to mention here that an analytic reduction through \texttt{FIRE6} was possible in a personal laptop (i7, 8-core, 16GB RAM) using the SDE approach which produced ${\cal O}(10^2)$ integrals for reduction in order to derive the total DE, while this was not possible using the standard approach which produced ${\cal O}(10^3)$ integrals (the analytic reduction was able only till sectors with 8 propagator).}. For some intermediate-sector (till 9 propagators) MI we used the \texttt{DlogBasis} \cite{3loop4} package combined with the SDE parametrization. More specifically, as we know \texttt{DlogBasis} in order to find FI of d-log form depends on the \textit{Spinor-Helicity Formalism}, which can not be applied when we deal with massive external momenta. When one deals with such problems the standard way to proceed is the decomposition of the external massive momentum in terms of two (arbitrary) massless momenta \cite{3loop4}, or the use of the Baikov representation \cite{Baik1,Baik2}. Another possible way of proceeding is the use of the SDE notation for the propagators where by definition the external momenta that appear in them for 1-mass problems are massless\footnote{The same approach can be used for 2-mass problems introducing an extra $y$ parameter beyond $x$ in order to catch the off-shellness of both masses.} and thus the spinor helicity parametrization can be applied. Thus while the command \cite{3loop4}
\begin{equation*}
{\small
\texttt{SetParametrization}[\texttt{SpinorHelicityParametrization}[\{l_1,l_2,l_3\}, \{a, b, c\}, \{q_1, q_2, q_3\}]]},
\end{equation*}
doesn't work when one uses the standard notation for the propagators with massive external momenta, it correctly works when one uses the SDE notation for the propagators and massless external momenta
\begin{equation*}
{\small
\texttt{SetParametrization}[\texttt{SpinorHelicityParametrization}[\{k_1,k_2,k_3\}, \{a, b, c\}, \{p_1, p_2, p_3\}]]}.
\end{equation*}
The last but most used method that we applied for the completion of the UT basis is the \textit{Building-Blocks} method\cite{buildingblocks}. In our study apart from the standard approach of using one-loop UT MI (massless boxes, triangles and bubbles with up to three external massive particles) as building blocks we also used UT basis elements from the massless planar double-box families with up to three external off-shell legs \cite{3loop6,Baik2,Henn}.

In intermediate steps we checked that the chosen basis elements were indeed UT by semi-numerical (keeping only $x$ analytic) derivations of the DE. For sectors with multiple MI where it is difficult to understand which of the chosen basis elements is not UT, a hint was given to us by the \texttt{C++} version of \texttt{Fuchsia} \cite{fuchsia}.

\section{Conclusion}

Within this contribution we presented the application of the SDE approach for the computation of the massless ladder-box families with up to one external off-shell, which have been previously solved in the literature using the standard DE approach. We also briefly discussed the methods that we used in order to obtain a DE of canonical form for the massless tennis-court families with one external off-shell leg. These UT basis will be made available together with the solutions of these families in a forthcoming publication.

At the moment we are working on the computation of the boundary conditions of the tennis-court families. In fact, we have already solved the F2 family and we have developed some new tools for the computation of boundaries within the SDE approach which are also applicable in the F3 family. Moreover in order for our results to be available for fast evaluations in phenomenological applications, we are currently working on the analytic continuation of the solutions to the three physical regions
\begin{equation*}
\begin{split}
&{\color{red}\text{1)}} \, m^2>0, \, \, \, \, s \geq m^2, \, \, \, \, t\leq 0, \, \, \, \, u \leq 0\\
&{\color{red}\text{2)}} \, m^2>0, \, \, \, \, s\leq 0, \, \, \, \, t\geq m^2, \, \, \, \, u \leq 0\\
&{\color{red}\text{3)}} \, m^2>0, \, \, \, \, s\leq 0, \, \, \, \, t\leq 0, \, \, \, \, u \geq m^2 \,.
\end{split}
\end{equation*}
of this scattering process.

As future work, encouraged by the efficiency of the SDE approach in dealing with three-loop problems and the phenomenological applications of these problems, we are planning to study the non-planar three-loop four-point massless families with one off-shell leg (15 families), starting from the non-planar ladder-boxes (4 families).


\section*{Acknowledgements}
We want to thank the organizers of {\it 15th International Symposium on Radiative Corrections: Applications of Quantum Field Theory to Phenomenology AND LoopFest XIX: Workshop on Radiative Corrections for the LHC and Future Colliders (Radcor and LoopFest 2021)} for
the very interesting conference.

\paragraph{Funding information}
The research work of D.C. was supported by the Hellenic Foundation for Research and Innovation (HFRI) under the HFRI Ph.D. Fellowship grant (Fellowship Number: 554).




\nolinenumbers

\end{document}